\documentclass[twocolumn]{article}  
\usepackage{graphicx}
\usepackage{txfonts}
\usepackage{natbib}
\begin{document}
\title{Multimessenger search for point sources: ultra-high energy cosmic rays and neutrinos}


\author{Jelena Petrovic, Nikhef, The Netherlands
}



 
\maketitle
The origin of ultra-high energy cosmic rays (UHECRs) and neutrinos is still a mystery. 
Hadronic acceleration theory suggests that they should originate in the same sources (astrophysical or cosmological),
together with gamma-rays. While gamma-rays have been linked to astrophysical sources, no point
source of UHECRs or neutrinos have been found so far. 
In this paper, the multimessenger combination of UHECRs and neutrinos as a 
new approach to the high energy particle point source search is suggested.
A statistical method for cross-correlation of UHECR and neutrino data sets is proposed.
By obtaining the probability density function of number of neutrino events within chosen 
angular distance from observed UHECRs, the
number of neutrino events in the vicinity of observed UHECRs, 
necessary to claim a discovery with a chosen significance, can be calculated. Different angular distances (bin sizes) are
considered due to the unknown deflection of cosmic rays in galactic and intergalactic magnetic fields.
Possible observed correlation of the arrival directions of UHECRs and neutrinos would provide 
a strong indication of hadronic acceleration theory. Correlation of both types of messengers with the location
of certain sets of observed astrophysical objects would indicate sites of acceleration. Any systematic offset 
in arrival directions between UHECRs and neutrinos may shed more light on magnetic field deflection of cosmic rays.

\section{Introduction}

The origin of high energy particles coming from the Universe is still not 
known. Currently, it is suggested that cosmic rays (protons and nuclei) up to about 10$^{15}$eV 
gain their energy in acceleration by shocks from supernova explosions 
\citep[see for example][]{2000RvMP...72..689N}.
For cosmic rays with energies around 10$^{16-17}$eV, it has been suggested that acceleration can occur in 
the interaction of particles with multiple supernova remnants \citep{1991aame.conf..273I}, galactic wind 
\citep{2004A&A...417..807V}, and microquasars \citep{2008arXiv0805.2378M}. 
The highest energy charged cosmic rays are expected to be of extragalactic origin and get accelerated in
jets of gamma-ray bursts \citep{1995PhRvL..75..386W,1995ApJ...453..883V} or
active galactic nuclei \citep{1987ApJ...322..643B,1993A&A...272..161R}. 
Another option are so-called top-down scenarios:
protons and neutrons, but also neutrinos and gamma-rays, are produced from quark and gluon 
fragmentation of heavy exotic particles formed in the early Universe
\citep{hill,1983ICRC....2..393S}.
According to the theory of hadronic acceleration, the ultra-high energy 
cosmic ray (UHECR) flux is expected to be accompanied
by associated fluxes of gamma-rays and neutrinos
from pion decays formed in the collision of protons with photons  
\citep[for example see][]{2005NJPh....7..130Z}.

When a cosmic ray interacts with a particle in the Earth's atmosphere, 
a shower of particles is produced propagating towards the ground with almost the 
speed of light. These so-called "cosmic ray air showers" can be detected by arrays of 
Cherenkov detectors, fluorescent detectors or radio antennas. 
Low energy cosmic rays are constantly bombarding the Earth producing 
such showers, but cosmic rays of energies above 10$^{19}$eV are very rare.  
They are expected to occur less than once a year on an area of one 
square kilometer and large shower arrays like the Pierre Auger Observatory (3000km$^2$)
are needed.
Moreover, cosmic rays are deflected by magnetic fields on their way from source to the 
Earth, thus their arrival directions do not point back to their sources. 
The expected deflection due to the Galactic magnetic field is in the order of few degrees for 
ultra-high energy protons \citep{2005APh....24...32T}.  
Also, it is not yet known whether large deflections are caused by extra-galactic magnetic fields.
According to some authors \citep{2005JCAP...01..009D} they are in the order of one degree and bellow, 
but some other analysis are predicting far larger numbers \citep{2004PhRvD..70d3007S}. 
The highest energy cosmic rays are expected to be suppressed by interaction with the cosmic
microwave background radiation, which is known as the Greisen-Zatsepin-Kuzmin (GZK) cut-off
\citep{greisen,zatsepin}. Due to this effect, they can be observed only if they originate in sources 
closer than about 200Mpc. Note that the size of the GZK sphere is still under discussion, and
significantly smaller GZK spheres are suggested by some authors 
\citep[see for example][]{2008PhR...458..173B}. 
The HiRes experiment \citep{2004PhRvL..92o1101A} measurements are in agreement with the 
predicted flux suppression of 
the highest energy cosmic rays, however the results of the AGASA experiment 
\citep{1998PhRvL..81.1163T} show an excess
compared to the predicted GZK suppression \citep{2003PhLB..556....1B}.
The Pierre Auger collaboration reported recently observations consistent
with GZK cut-off \citep{2008PhRvL.101f1101A}.

Neutrinos are difficult to detect due to their small cross section for interaction with matter.
They can interact within the Earth's atmosphere and trigger air showers. The probability for interaction 
increases with the path length, thus they are most likely to
produce very inclined cosmic ray air showers. Those showers may be triggered anywhere along the 
traveling path and also close to the ground
\citep{1998APh.....8..321C}. However, a neutrino is much more likely to undergo charged current 
interaction with nuclei in the Earth's interior, producing a muon that emits Cherenkov 
radiation while moving through water or ice faster than the speed of light in the
corresponding medium. This principle is used in current 
neutrino telescopes, such as IceCube and ANTARES. 
The IceCube neutrino telescope is currently being built at the South Pole. 
The final configuration should reach around 80
strings with photomultiplier tubes (PMTs) in a 1km$^3$
geometric volume. Currently, data are being taken with 40 strings. 
ANTARES was recently finished in the Mediterranean
Sea close to the French coast. Its final configuration consists of 12 strings with PMTs
in 0.03km$^3$ of instrumented volume. 
The largest problem for observation of cosmic neutrinos 
is the large atmospheric flux of secondary 
neutrinos coming from air showers. 
Comparing to UHECRs, the advantage of neutrinos as messengers 
is that they are not deflected in magnetic fields, so their arrival direction 
point back to their sources. 
The neutrino interaction cross section increases with increasing energy, 
so the highest energy neutrinos are more likely to interact with the detector, but also within
the Earth's interior. This means that neutrinos above around 1PeV can be observed only if coming from  
horizontal directions, so their path through the Earth interior is shorter.



\begin{figure}
\begin{center}
\includegraphics*[width=\columnwidth,clip]{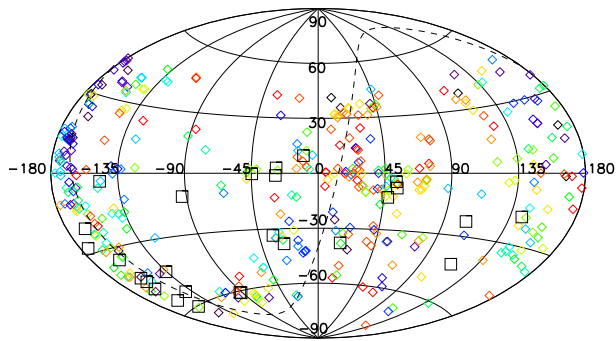}
\caption{\label{auger}Aitoff skyplot of 27 UHECR events with E$>$57EeV (squares) observed by the Pierre Auger
Observatory and AGNs from the VCV catalog at D$<$75Mpc
(diamonds) in equatorial coordinates. 
The color code represents the redshift of AGNs, with purple being the closest and red
being the furthest objects.  The dashed line
represent the supergalactic plane (plot after \citet{2008APh....29..188P}) }
\end{center}
\end{figure}
\begin{figure*}[t]
\begin{center}
\includegraphics*[width=0.9\columnwidth,clip]{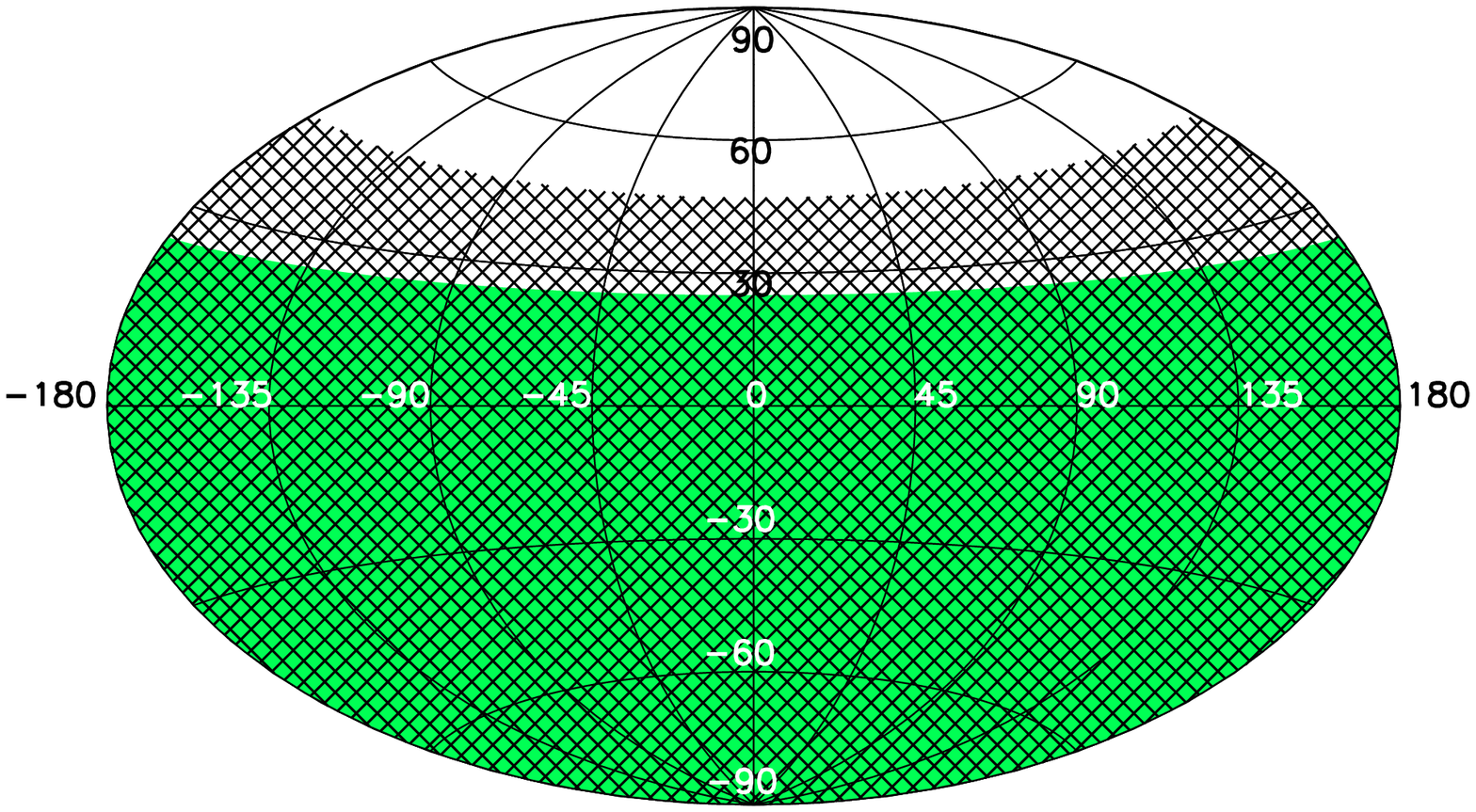}
\includegraphics*[width=0.9\columnwidth,clip]{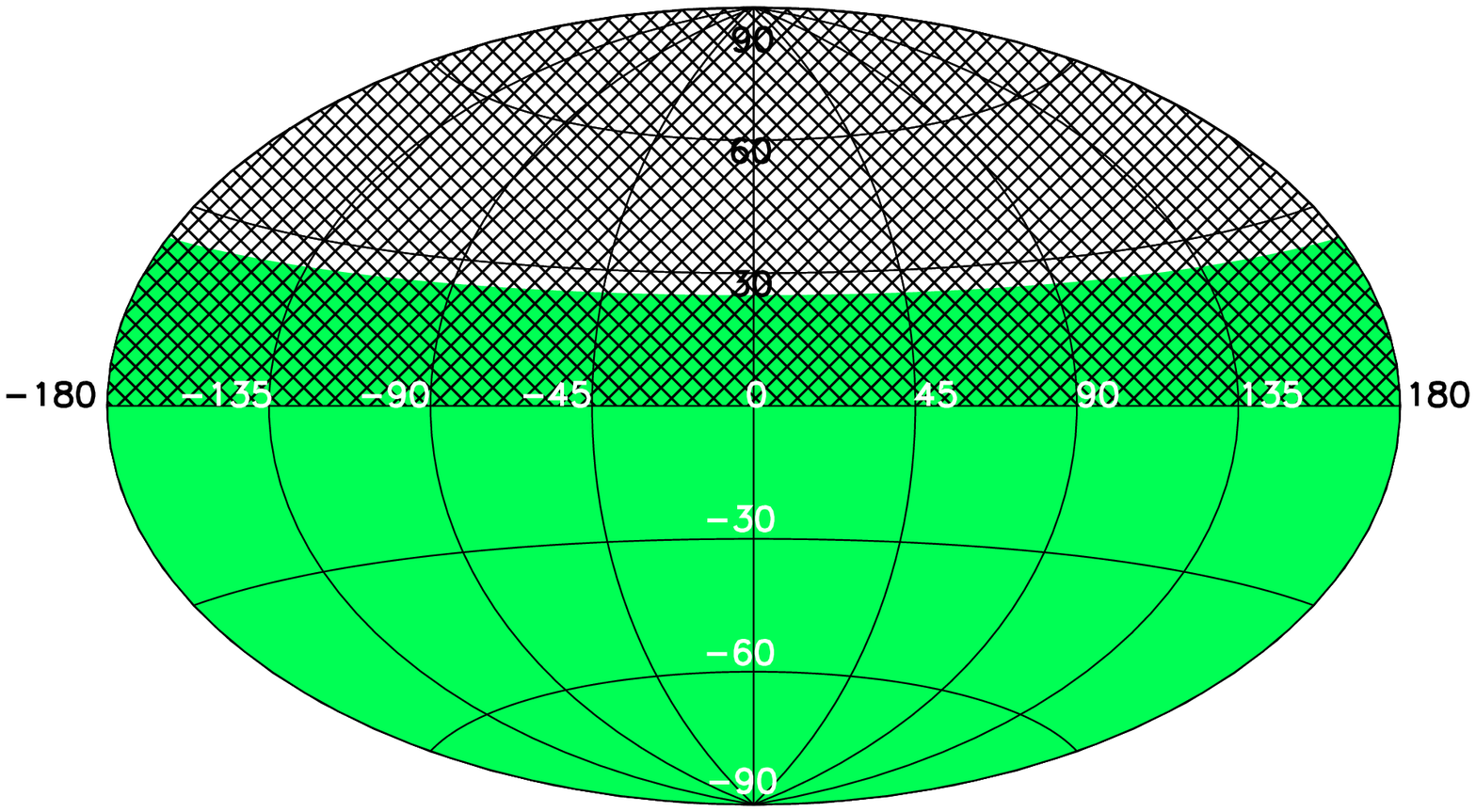}
\caption{\label{view}Aitoff maps of combined sky visibility in equatorial coordinates. 
With the additional cut on zenith angles for UHECR observations, 
Z$<$60$^{\circ}$, being taken into account, the field of view for the Pierre Auger Observatory 
is beneath declinations of 25$^{\circ}$ (on both panels: green).
The left panel represents additionally the ANTARES neutrino telescope field of view (hatched).
Visible declinations are beneath 47$^{\circ}$. 
The right panel shows additionally the IceCube neutrino telescope field of view (hatched). Visible declinations (for up-going
events) are above 0$^{\circ}$.}
\end{center}
\end{figure*}

\section{Cross-correlation of neutrinos and cosmic rays}

The Pierre Auger Observatory reported an anisotropy in the arrival directions of ultra high energy cosmic rays 
\citep{2008APh....29..188P}.
Correlation with Active Galactic Nuclei (AGN) from the VCV catalog \citep{2006A&A...455..773V} 
was the most significant for cosmic rays with energies higher than 
57EeV and AGNs at distances less than $~$75Mpc.
For this correlation, 
an angular separation between AGNs and cosmic rays larger than the point spread function of the detector is
considered, so magnetic deflection of cosmic rays is in this way partially taken into account. 
However, the mass composition of 
the observed events is not yet established,
and the possible deflection in magnetic fields is not yet known. This means that in the case that the
observed events
are mostly nuclei, their arrival directions may significantly differ
from the directions of their acceleration sites. The anisotropy was estimated with a confidence 
level of 99\%.
The suggested correlation with local AGN sources 
mostly following the location of the supergalactic plane decreased in the following analysis 
\citep{2009arXiv0906.2347T} with 58 events observed at energies above 55EeV. 
The anisotropy of observed events remains with a confidence 
level of 99\%. 
On the other side, no significant excess above the atmospheric neutrino flux was reported by
the neutrino telescopes IceCube and ANTARES \citep{2009arXiv0905.2253I,toscano}.

Instead of searching for such a localized excess, neutrino arrival direction can be correlated
with the arrival directions of ultra high energy cosmic rays.
In this paper, a method is suggested for such a correlation. 
The described analysis is based on the calculation of the probability density function of the number of neutrino events
within specific angular distances from UHECRs. 
For a given set of observed cosmic ray arrival directions that are used as a catalog, Monte Carlo 
neutrino sets are generated, and angular distances between all cosmic rays and all MC neutrinos are calculated, 
(full sky correlation). 
In this way we obtain neutrino counts in chosen bins around the observed UHECRs, expected 
only from the atmospheric background neutrino flux.
In other words, we calculate the probability that any neutrino count appears in the observed data.
Usually, the bin size for this kind
of analysis is the angular resolution of the detector, but due to the unknown 
deflection of cosmic rays in magnetic fields, we leave this parameter free.
This allows our analysis to be sensitive for a potential systematic offset between arrival directions of
ultra-high energy cosmic rays and neutrinos.

An example for the calculation of the probability density function of the number of neutrino events
within specific angular distances from UHECRs is presented. Fixed 27 UHECRs events
are used as catalog sources. 
Monte Carlo simulations with 1000 neutrinos on each generated sky map,
randomly distributed over the full sky (declinations [90$^{\circ}$,-90$^{\circ}$] and right ascension 
[0$^{\circ}$,360$^{\circ}$]) are 
performed. Further, angular distances between all Monte Carlo neutrinos 
and 27 UHECRs are calculated, for each produced sky map. This provides the expected angular distance 
distribution between neutrinos and cosmic rays in the absence of any 
correlation, over the whole sky.
For a specific neutrino telescope, such as ANTARES or IceCube, 
instead of the whole sky, only declinations visible by the chosen detector should be used, and 
instead of a completely random neutrino background distribution, 
one that corresponds to the detector exposure should be applied.
Also, the actually observed number of neutrino events should be used for Monte Carlo calculations.
Correlation should be performed on a part of sky visible for both UHECR and neutrino detector.
Fig. \ref{view} shows the combined fields of view for the Auger observatory and the ANTARES telescope (left panel)
and the Auger observatory and the IceCube telescope (right panel).
The Auger observatory is located at 35$^{\circ}$S, and declinations up to 55$^{\circ}$ are observable. 
However, for the observed UHECR, an additional cut on zenith angle of 60$^{\circ}$ is added,
which means that declinations up to about 25$^{\circ}$ are visible. The ANTARES telescope is located at 43$^{\circ}$N,
and for up-going events, declinations up to 47$^{\circ}$ are visible. The IceCube detector is at the South pole, so 
everything above 0$^{\circ}$ declination is in the field of view. IceCube is also investigating the possibility of 
extended sky search,
by using also down-going events, and in that case all declinations above around -50$^{\circ}$ are observable
\citep{2009arXiv0903.5434L}.

\begin{figure}
\begin{center}
\includegraphics*[width=\columnwidth,clip]{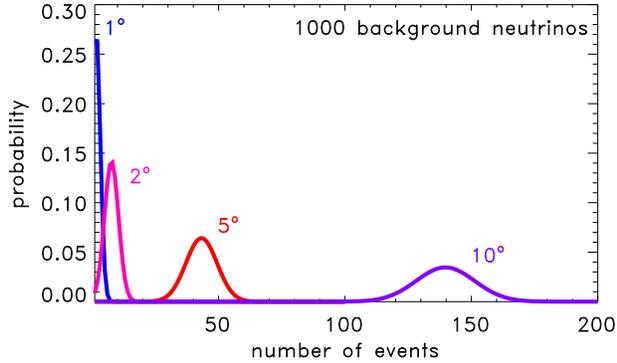}
\caption{\label{dist}Probability density function of the number of neutrino events on 
1$^{\circ}$,2$^{\circ}$,5$^{\circ}$ and 10$^{\circ}$ angular distances from 
27 fixed UHECR, for 1000 random background neutrinos.}
\end{center}
\end{figure}

Using this probability density function, the
expected number of neutrino events in the vicinity of the observed ultra-high energy cosmic rays, 
necessary to claim a discovery with a chosen significance, can be calculated.
Even though the expected angular resolution for IceCube and ANTARES is better than 1 degree, bin sizes 
of 1-10 degrees are considered. This is because UHECRs may be deflected significantly in magnetic
fields on their path to the Earth, thus the analysis is sensitive to a possible offset between
arrival directions of cosmic rays and neutrinos. 

Probabilities for the expected number of background neutrinos, in bins around the 27 fixed UHECR events are shown in 
Fig. \ref{dist}.
For example, the probability to observe 1 neutrino,
within 27 bins of 1$^{\circ}$ is around 26\%, 
and the probability to observe 5 
neutrinos within those bins is around 3\%. 
The probability to not have any neutrinos is around 14\%, 
and to have more than 8 is close to zero. The same distributions are shown
for bin sizes of 2,5 and 10 degrees.
This can be translated into 
discovery potential and
the number of neutrino events within a certain distance from observed cosmic rays, necessary
to claim a discovery, can be obtained.

\begin{figure*}[t]
\begin{center}
\includegraphics*[width=\columnwidth,clip]{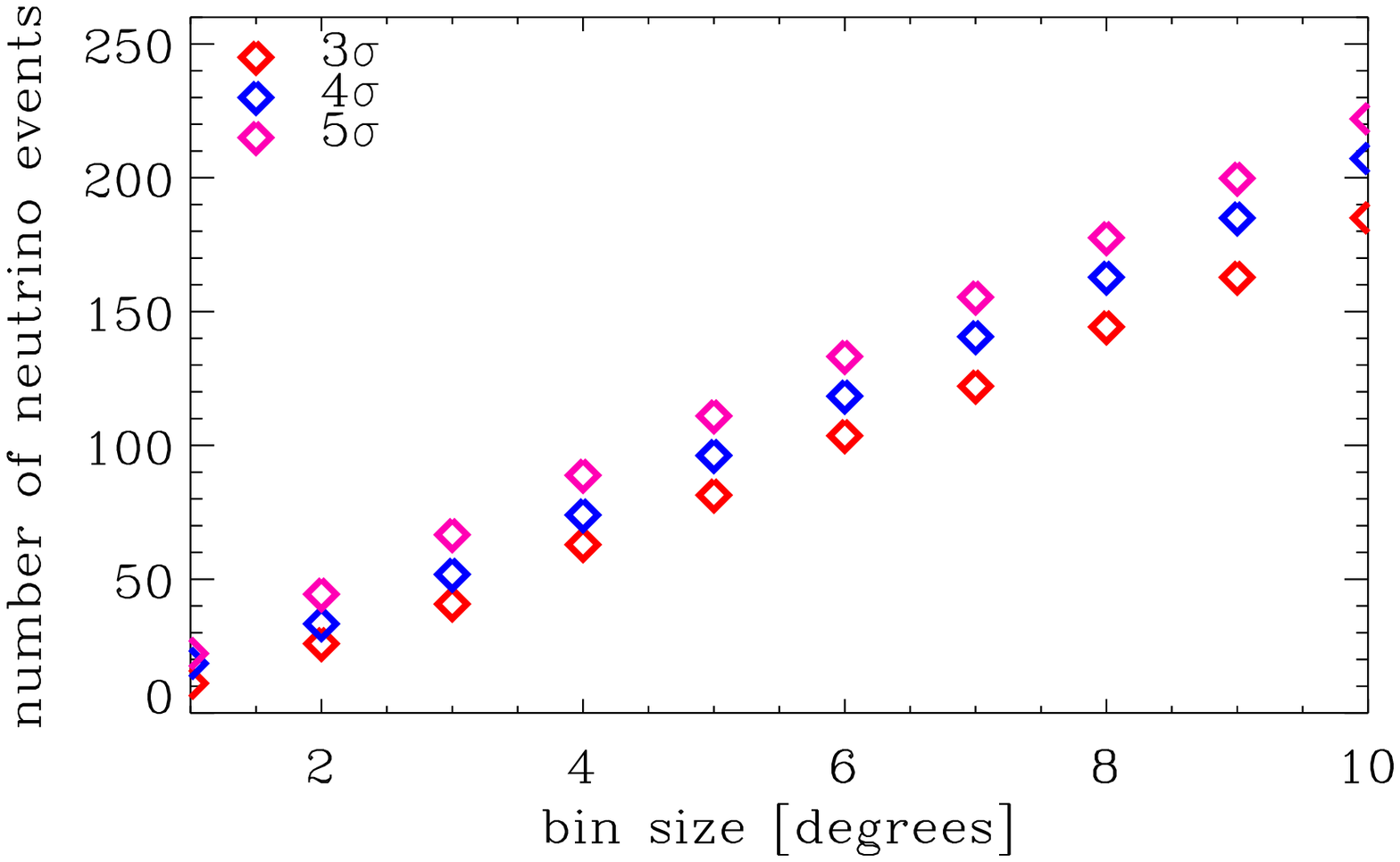}
\includegraphics*[width=\columnwidth,clip]{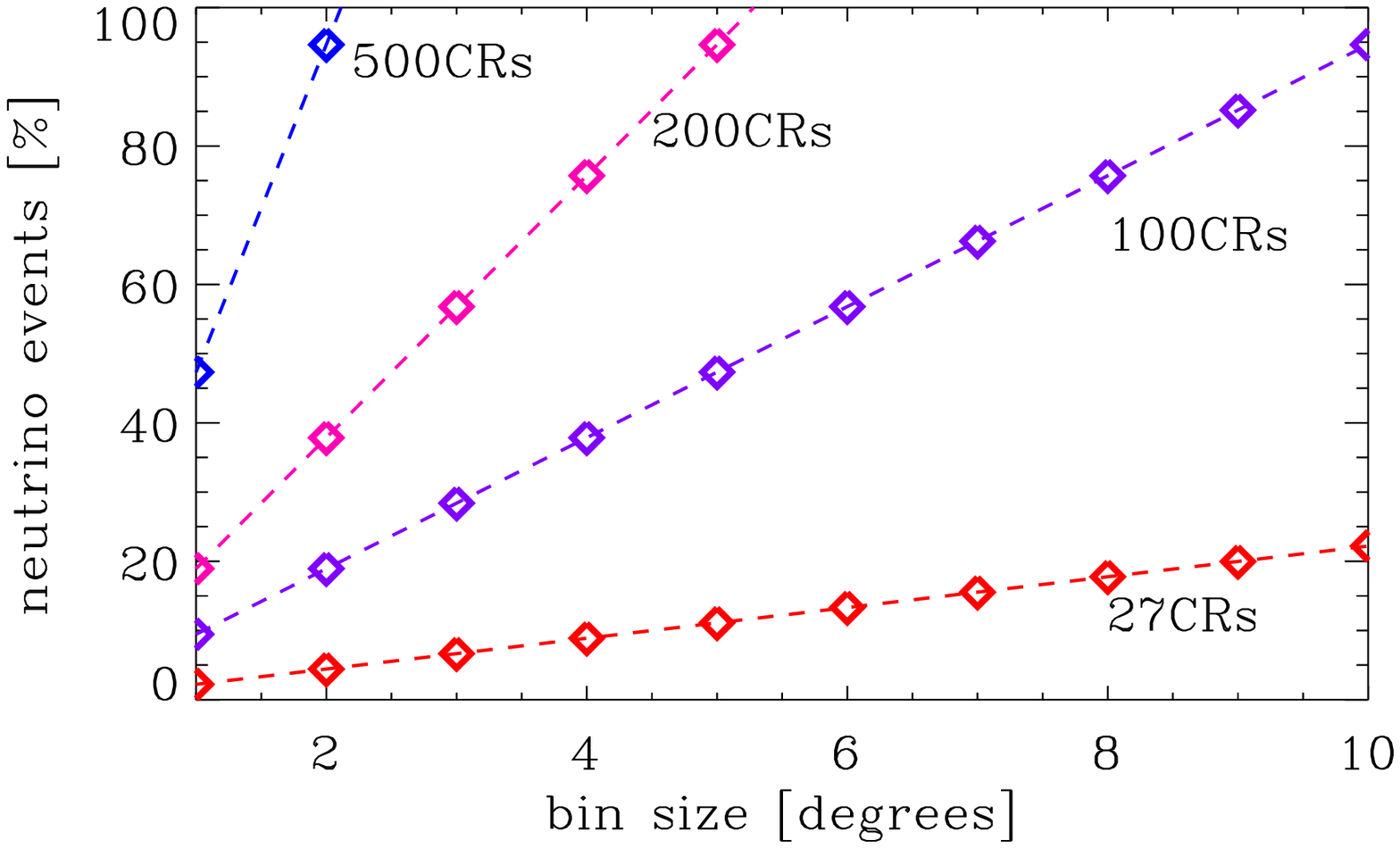}
\caption{\label{sigma}Left panel: The number of neutrino events 
necessary to claim 3$\sigma$ (red), 4$\sigma$ (blue) and 5$\sigma$ (magenta) correlation
significance, as a function of the bin size, for the example of 27 fixed UHECRs and 1000 random neutrinos.
Right panel: The percentage of neutrino events necessary to claim 5$\sigma$ correlation significance,
as a function of the bin size and number of UHECRs.}
\end{center}
\end{figure*}

Depending on Monte Carlo statistics, the number of neutrino events in the vicinity of cosmic rays, 
needed for a 3, 4 or 5$\sigma$ discovery (correlation) claims can be established. 
This is shown in the left panel of Fig. \ref{sigma} for 10$^7$ Monte Carlo
simulations, 27 fixed UHECRs, 1000 random neutrinos, and bin sizes of 1-10$^{\circ}$.
For example, correlation significance of 3$\sigma$ is reached with 
about 11 neutrino events within 27 bins of 1$^{\circ}$, 4$\sigma$ with 18 and 5$\sigma$ with 22.
The right panel of Fig. \ref{sigma} shows the percentage of neutrino events necessary for 5$\sigma$
discovery claim, depending on the bin size and the number of UHECRs. The bottom red line corresponds to
magenta values presented in the left panel of Fig. \ref{sigma}. 
It is shown that in the case of 27 UHECRs, correlation significance of 
5$\sigma$ is reached with 2\%-25\% of neutrino
events falling in 1-10$^{\circ}$ bins. However, with 100 UHECRs, it becomes more difficult to claim 
5$\sigma$ discovery for large bins, since 50-100\% of neutrino events should be in 5-10$^{\circ}$ bins.
This becomes even more extreme for 200 and 500 UHECRs, where more than 80\% of neutrino events should
be already in 4$^{\circ}$ and 2$^{\circ}$ respectively, to claim a 5$\sigma$ correlation. 
This stays
the case even if a far larger amount of neutrinos is added (50000). However, it does slightly 
depend on UHECRs sky distribution, since the exposure for neutrino detection depends on declination. 
Anyhow, in the case of large number ($>$100) of 
observed UHECRs, it is suggested to first group those in clusters, and then apply the described 
method on centers of those UHECR clusters.

Instead of the full (visible) sky, the UHECR-neutrino cross correlation can also be applied only on certain regions.
The most interesting regions, according to a current Auger results, are Centaurus A and the supergalactic plane due to
the observed event excess. Also, the Virgo cluster from where no UHECR was yet observed, should be considered,
since it might happen that due to a large magnetic field deflection, events are moved to other arrival
directions.

\begin{figure*}[t]
\begin{center}
\includegraphics*[width=\columnwidth,clip]{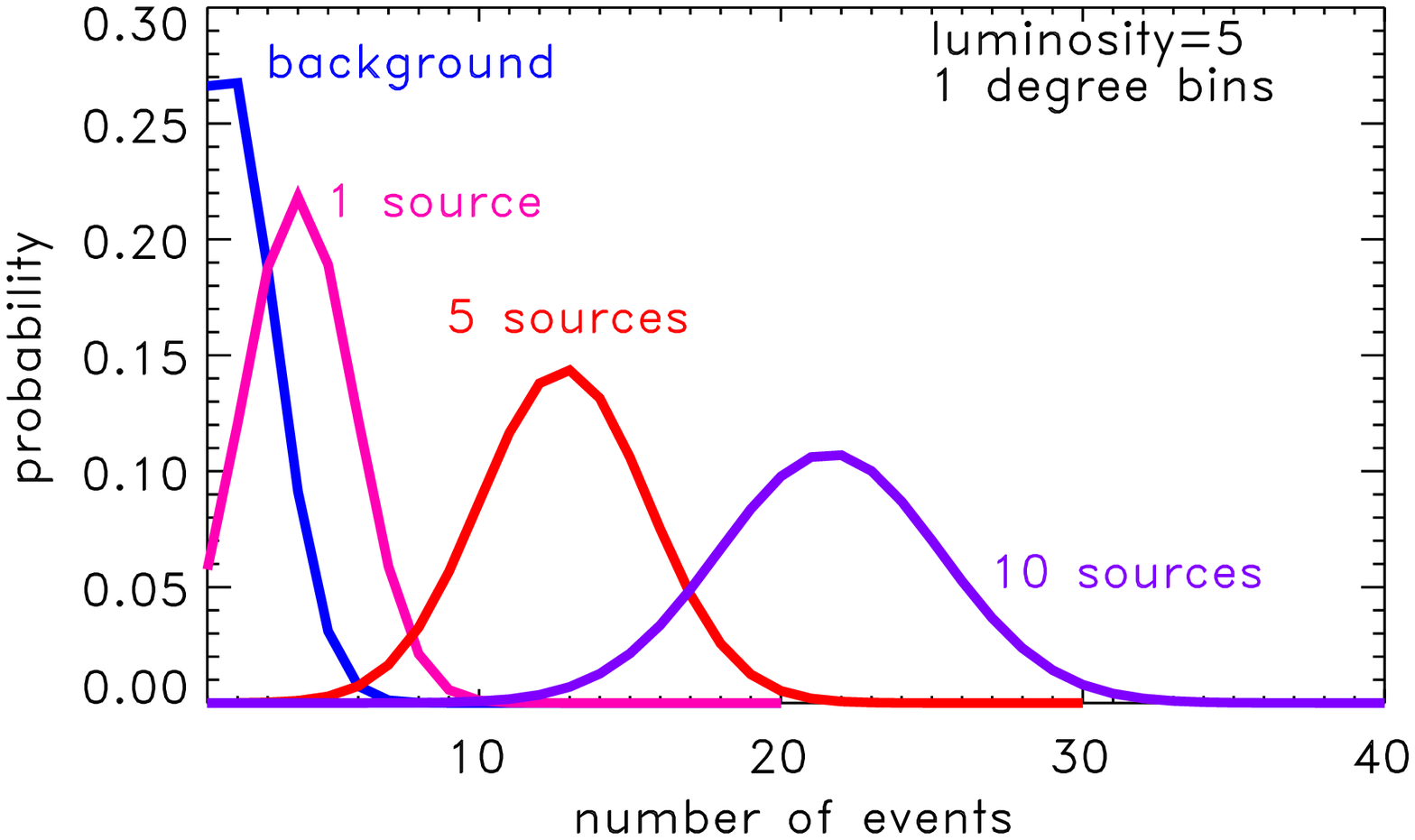}
\includegraphics*[width=\columnwidth,clip]{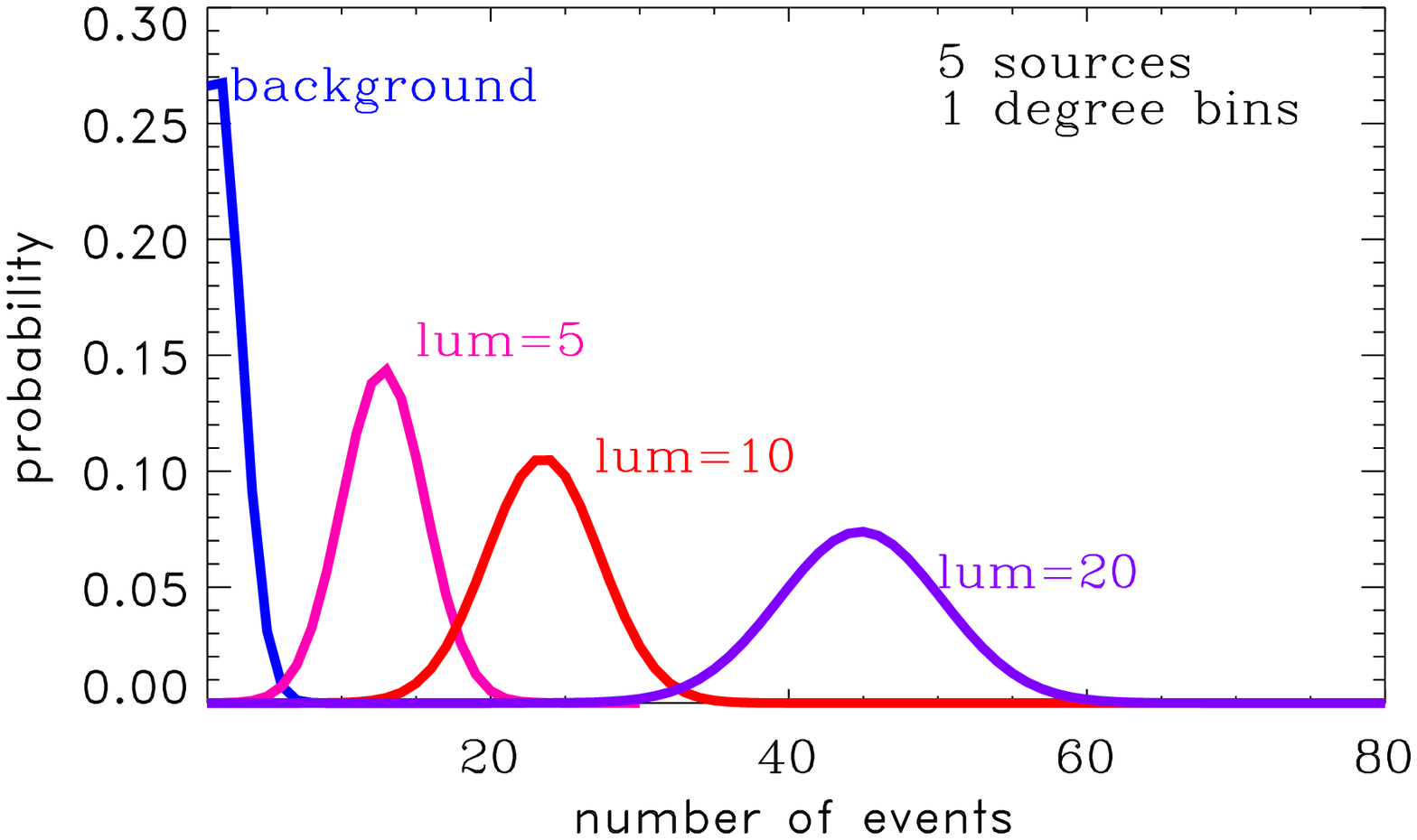}
\caption{\label{lum5}Left panel: Probability density function of the number of neutrino events 
within 1$^{\circ}$ angular distance
from 27 fixed UHECRs, for 
1000 random neutrinos + fake neutrino sources (1,5 and 10) coinciding with UHECRs.
Right panel: 
Probability density function of the number of neutrino events 
within 1$^{\circ}$ angular distance
from 27 fixed UHECRs, for 
1000 random neutrinos + 5 fake neutrino sources with luminosities 5, 10 and 20 neutrinos, 
coinciding with fixed UHECRs.
}
\end{center}
\end{figure*}

To test the method, fake neutrino sources that coincide with 
UHECR events are added to the Monte Carlo 
simulations. In this case, the expected number of neutrinos increases, 
proportional to the assumed number of neutrino sources coinciding with fixed UHECRs.
This is shown in the left panel of Fig. \ref{lum5}. 
The luminosity of all fake sources is assumed to be 5 neutrinos 
(Gaussian-distributed around the given arrival direction of a cosmic ray),
and the bin size is taken to be 1 degree. 
For example, the probability of having 5 neutrinos within 27 bins with one fake source 
is around 20\%, compared to 5\% with only background neutrinos.
With 5 fake sources, the highest probability is to observe around 10 
neutrinos in 27 bins. The probability to observe more than 20 neutrinos is beneath 1\%.
With 10 fake sources, around 20 neutrinos are 
most likely to be observed, and more than 30 are expected in less than 1\% of the cases. 
The right panel of Fig. \ref{lum5} shows the expected distributions for 5 fake neutrino sources,
with luminosities 5, and exaggerated values of 10 and 20 neutrinos. 
If the added sources have higher luminosity, the expected number of neutrinos also
increases. In other words, as expected, the method recognizes a stronger correlation, 
when the neutrino sources coincide with the observed cosmic rays and when the neutrino sources are brighter.

\section{Discussion}

 \begin{figure*}[t]
\begin{center}
\includegraphics*[width=\columnwidth,clip]{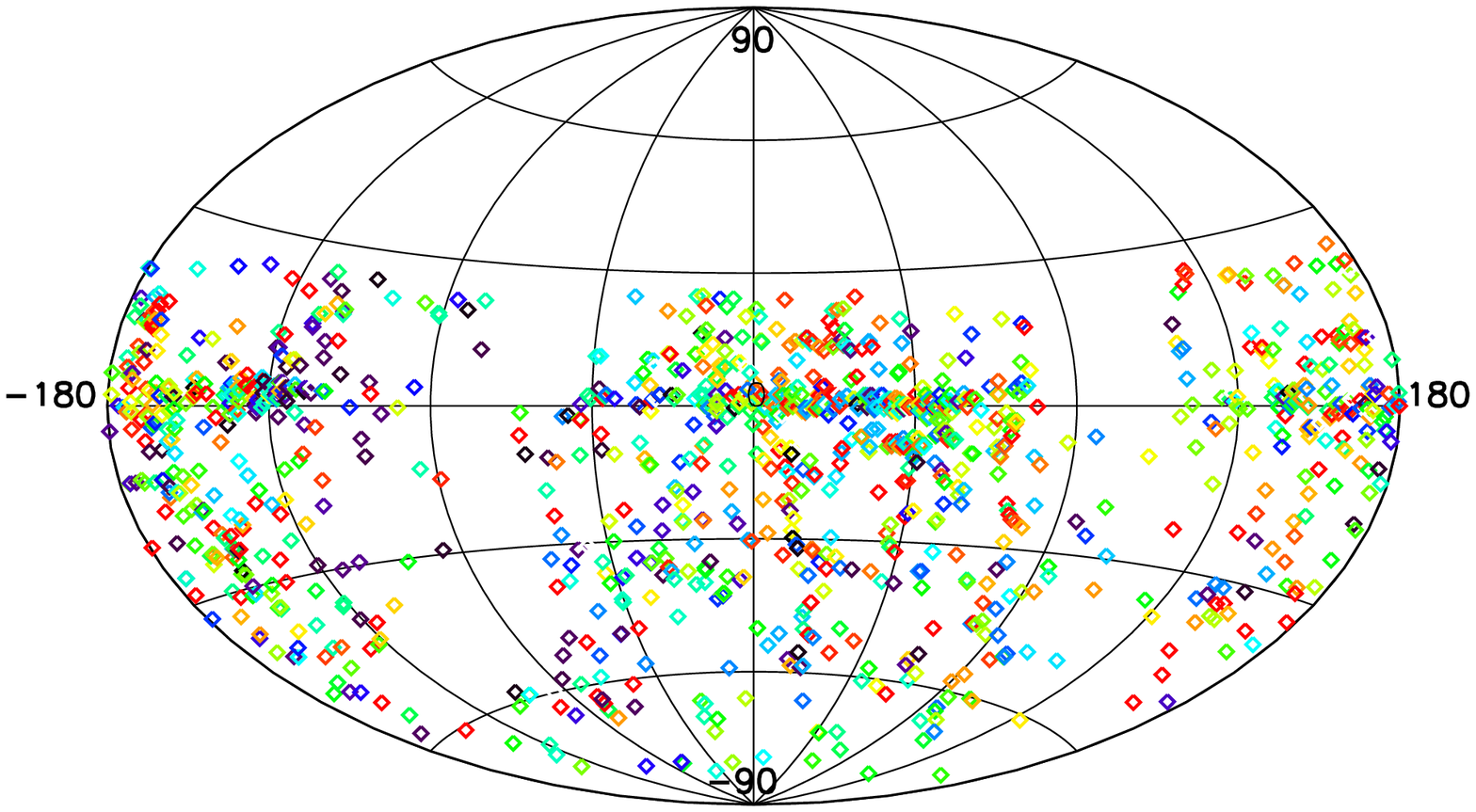}
\includegraphics*[width=\columnwidth,clip]{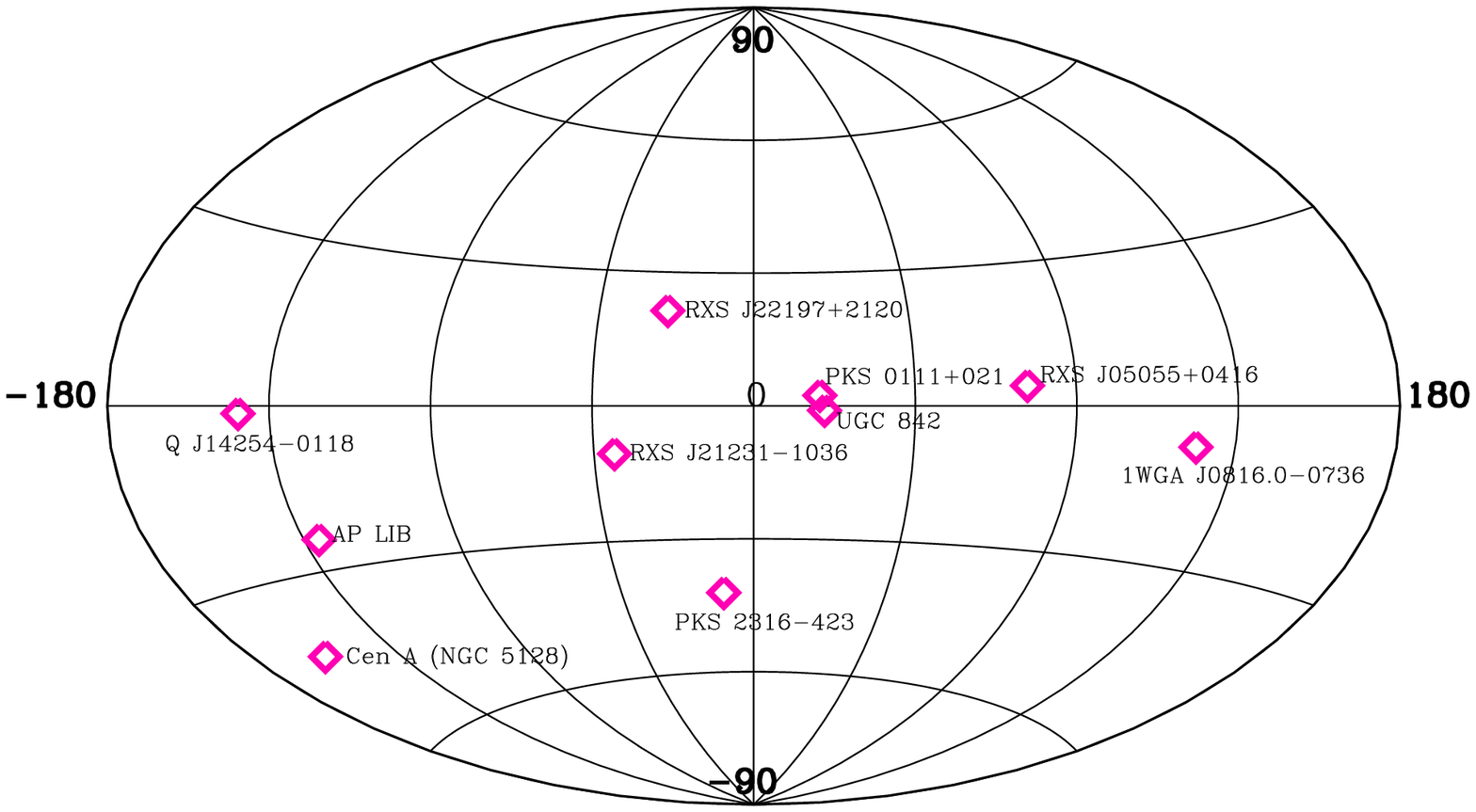}
\caption{\label{agn}Left panel: Aitoff skymap in equatorial coordinates of 1205 AGNs from VCV catalog within 
200Mpc (z$<=0.05$) visible for the Pierre Auger Observatory. 
Colors are representing the
redshift, with purple being the closest and red being the furthest objects. Right panel: Aitoff skymap in equatorial coordinates
of BL Lac objects (confirmed, probable and questionable) + Cen A
from VCV catalog within 200Mpc, visible for the Pierre Auger observatory.}
\end{center}
\end{figure*}

In the previous section, a statistical method for correlation of UHECR and neutrino
data sets was described. 
Possible observed correlation of the arrival directions of those two types of messengers would provide 
a strong indication of hadronic acceleration theory. Correlation of both types of messengers with the location
of certain sets of observed astrophysical objects would indicate sites of acceleration. 
As we already mentioned, it is currently believed that UHECR are originating in extragalactic sources,  
gamma-ray burst (GRBs) and active galactic nuclei (AGNs).
While gamma-ray bursts (GRBs) are transient extragalactic sources, 
and more suitable for correlation between gamma-rays and neutrinos (due to a significant
delay in the arrival time between protons/nuclei and neutrinos),
active galactic nuclei (AGNs), as extragalactic and steady sources, seem like a natural choice to explore the 
link between UHECRs and neutrinos. Assuming that there is a GZK cut-off, for an interpretation of correlation 
between arrival directions of UHECR and neutrinos,
the limit should be set on AGNs within the GZK sphere.
(left panel of Fig. \ref{agn}).
However, we should keep on mind that AGNs with jets not pointing towards the Earth are possible sources of UHECRs, 
since they can escape on the sides of jets, but neutrinos follow the direction of the jet 
and are possible observed only from blazar/BL Lac objects, or ones so close by that they show
blazar-like radiation, like it is the case with Centaurus A (right panel of Fig. \ref{agn}).
Due to UHECR excess observed in Cen A region by the Pierre Auger observatory, this region is interesting for UHECR/neutrino
link investigation. Neutrino fluxes from Cen A are recently calculated by \citet{2008PhRvD..78h3009K}
and \citet{2009APh....31..138B}, but if neutrinos are indeed following the direction of jets, neutrino telescopes might not see 
the signal from this source.

Any systematic offset in arrival directions between UHECRs and neutrinos may shed more light on 
magnetic field deflection of cosmic rays.
The Galactic magnetic field has been shown to have both regular and random components.
The regular component is believed to have a spiral structure and the corresponding global model
was done by \citet{1997ApJ...479..290S}. According to this model, protons with an energy of about 4x10$^{19}$eV
can be deflected by the regular galactic magnetic field around 5$^{\circ}$. \citet{2005APh....24...32T} calculated
deflection in the random component to be about 0.2$^{\circ}$-1.5$^{\circ}$ depending on the arrival 
direction of a 4x10$^{19}$eV proton. \citet{2005JCAP...01..009D} 
used simulations of the large-scale structure to investigate 
magnetic fields in the intergalactic medium. 
They constructed full sky-maps of expected deflections for protons with energies of 10$^{20}$eV and 
4x10$^{19}$eV, and found that strong deflections are only present if UHE protons 
cross a galaxy cluster, and even then stay beneath 1$^{\circ}$. On the other side, \citet{2004PhRvD..70d3007S} suggested
that deflection in extra-galactic magnetic fields can be of order 20$^{\circ}$ up to 10$^{20}$eV. 
They suggested that a region of strong magnetic field surrounding the observer would shield off
UHECRs from sources outside of the magnetized region, so the observed flux would be dominated
by a few closer sources and appear more anisotropic.

\section{Conclusions}
The hadronic acceleration theory predicts that ultra-high energy cosmic rays are accompanied
by neutrino and gamma-ray fluxes. While gamma-rays have been 
linked with astrophysical sources by observations of telescopes like H.E.S.S. and MAGIC, and several
attempts have been made to correlate them with neutrinos \citep{2007JPhCS..60..340H,2008ICRC....3.1257A,2008AN....329..334B},
the UHECR-neutrino connection has not yet been examined.
The reason is that both UHECRs and neutrinos have many down-sides as messengers.
UHECRs are rare and they do not point back to their sources, since they are scrambled by galactic and intergalactic 
magnetic fields. Also, due to the interaction with the cosmic microwave background photons, 
they might be limited to the distance of about 200Mpc or less. 
However, they are detectable with large shower arrays, like the Pierre Auger Observatory, which so far 
reported observation of few tens of events above 55EeV.
On the other side, cosmic neutrinos point back to their sources and their traveling distances are not limited, but they
interact rarely, which makes them difficult to detect. Also, they are 
for now lost in the huge atmospheric background of neutrinos coming from air showers, 
and so far none of the currently operating neutrino telescopes (IceCube, ANTARES) have 
observed neutrino excess above the atmospheric background flux.

In this paper, the multimessenger combination of UHECRs and neutrinos as a 
new approach to the high energy particle point source search is suggested.
A statistical method for cross-correlation of UHECR and neutrino data sets is proposed.
By obtaining the probability density function of the number of neutrino events within specific 
angular distance from observed UHECRs, the
number of neutrino events in the vicinity of observed ultra-high energy cosmic rays, 
necessary to claim a discovery with a chosen significance, can be calculated. Different bin sizes are
considered due to the unknown magnetic deflection of UHECRs.
This multimessenger connection is crucial to indicate that hadronic acceleration 
actually happens in astrophysical sources, since only then neutrinos are to be observed 
coming from those sources. 
Through the correlation of arrival directions of those two very different messengers,
we can gain additional insight into the non-thermal Universe. In the case of a correlation, we can 
learn more about the possible sources and physical processes that are responsible 
for the existence of ultra-high energy particles, such as hadronic 
acceleration of particles in sources. In the case of an offset between arrival directions of cosmic rays
and neutrinos, additional insight in UHECR deflection in magnetic fields can be gained.

I would especially like to thank Charles Timmermans for valuable comments. Many thanks also to Patrick Decowski and Corey Reed.
Additionally, I would like to thank the ANTARES and the Pierre Auger collaborations.
This work is supported by the NWO Veni grant of Jelena Petrovic.
 
\bibliographystyle{aa}

\bibliography{lela}

\begin{thebibliography}{34}
\expandafter\ifx\csname natexlab\endcsname\relax\def\natexlab#1{#1}\fi

\bibitem[{{Abbasi} {et~al.}(2004){Abbasi}, {Abu-Zayyad}, {Amann}, {Archbold},
  {Bellido}, {Belov}, {Belz}, {Bergman}, {Cao}, {Clay}, {Cooper}, {Dai},
  {Dawson}, {Everett}, {Fedorova}, {Girard}, {Gray}, {Hanlon}, {Hoffman},
  {Holzscheiter}, {H{\"u}ntemeyer}, {Jones}, {Jui}, {Kieda}, {Kim}, {Kirn},
  {Loh}, {Manago}, {Marek}, {Martens}, {Martin}, {Matthews}, {Matthews},
  {Meyer}, {Moore}, {Morrison}, {Moosman}, {Mumford}, {Munro}, {Painter},
  {Perera}, {Reil}, {Riehle}, {Roberts}, {Sarracino}, {Sasaki}, {Schnetzer},
  {Shen}, {Simpson}, {Sinnis}, {Smith}, {Sokolsky}, {Song}, {Springer},
  {Stokes}, {Taylor}, {Thomas}, {Thompson}, {Thomson}, {Tupa}, {Westerhoff},
  {Wiencke}, {Vanderveen}, {Zech}, \& {Zhang}}]{2004PhRvL..92o1101A}
{Abbasi}, R.~U., {Abu-Zayyad}, T., {Amann}, J.~F., {et~al.} 2004, Physical
  Review Letters, 92, 151101

\bibitem[{{Ackermann} {et~al.}(2008){Ackermann}, {Bernardini}, {Galante}, \&
  {et al.}}]{2008ICRC....3.1257A}
{Ackermann}, M., {Bernardini}, E., {Galante}, N., \& {et al.} 2008, in
  International Cosmic Ray Conference, Vol.~3, International Cosmic Ray
  Conference, 1257--1260

\bibitem[{{Bahcall} \& {Waxman}(2003)}]{2003PhLB..556....1B}
{Bahcall}, J.~N. \& {Waxman}, E. 2003, Physics Letters B, 556, 1

\bibitem[{{Becker}(2008)}]{2008PhR...458..173B}
{Becker}, J.~K. 2008, Phys.Rep., 458, 173

\bibitem[{{Becker} \& {Biermann}(2009)}]{2009APh....31..138B}
{Becker}, J.~K. \& {Biermann}, P.~L. 2009, Astroparticle Physics, 31, 138

\bibitem[{{Biermann} \& {Strittmatter}(1987)}]{1987ApJ...322..643B}
{Biermann}, P.~L. \& {Strittmatter}, P.~A. 1987, Apj, 322, 643

\bibitem[{{Bouwhuis}(2008)}]{2008AN....329..334B}
{Bouwhuis}, M. 2008, Astronomische Nachrichten, 329, 334

\bibitem[{{Capelle} {et~al.}(1998){Capelle}, {Cronin}, {Parente}, \&
  {Zas}}]{1998APh.....8..321C}
{Capelle}, K.~S., {Cronin}, J.~W., {Parente}, G., \& {Zas}, E. 1998,
  Astroparticle Physics, 8, 321

\bibitem[{{Dolag} {et~al.}(2005){Dolag}, {Grasso}, {Springel}, \&
  {Tkachev}}]{2005JCAP...01..009D}
{Dolag}, K., {Grasso}, D., {Springel}, V., \& {Tkachev}, I. 2005, Journal of
  Cosmology and Astro-Particle Physics, 1, 9

\bibitem[{{Greisen}(1966)}]{greisen}
{Greisen}, K. 1966, Physical Review Letters, 16, 748

\bibitem[{{Hill}(1983)}]{hill}
{Hill}, C.~T. 1983, Nucl.Phys.B, 224, 469

\bibitem[{{Hughey} \& {the Ice Cube Collaboration}(2007)}]{2007JPhCS..60..340H}
{Hughey}, B. \& {the Ice Cube Collaboration}. 2007, Journal of Physics
  Conference Series, 60, 340

\bibitem[{{IceCube Collaboration: R.~Abbasi}(2009)}]{2009arXiv0905.2253I}
{IceCube Collaboration: R.~Abbasi}. 2009, arxiv:astro-ph/0905.2253

\bibitem[{{Ip} \& {Axford}(1991)}]{1991aame.conf..273I}
{Ip}, W.-H. \& {Axford}, W.~I. 1991, in Astrophysical Aspects of the Most
  Energetic Cosmic Rays, 273

\bibitem[{{Koers} \& {Tinyakov}(2008)}]{2008PhRvD..78h3009K}
{Koers}, H.~B.~J. \& {Tinyakov}, P. 2008, Phys. Rev. D, 78, 083009

\bibitem[{{Lauer} \& {the IceCube Collaboration}(2009)}]{2009arXiv0903.5434L}
{Lauer}, R. \& {the IceCube Collaboration}. 2009, arxiv:astro-ph/0903.5434

\bibitem[{{Mirabel}(2008)}]{2008arXiv0805.2378M}
{Mirabel}, I.~F. 2008, arxiv:astro-ph/0805.2378

\bibitem[{{Nagano} \& {Watson}(2000)}]{2000RvMP...72..689N}
{Nagano}, M. \& {Watson}, A.~A. 2000, Reviews of Modern Physics, 72, 689

\bibitem[{{Rachen} \& {Biermann}(1993)}]{1993A&A...272..161R}
{Rachen}, J.~P. \& {Biermann}, P.~L. 1993, A\&A, 272, 161

\bibitem[{{Schramm} \& {Hill}(1983)}]{1983ICRC....2..393S}
{Schramm}, D.~N. \& {Hill}, C.~T. 1983, in International Cosmic Ray Conference,
  393

\bibitem[{{Sigl} {et~al.}(2004){Sigl}, {Miniati}, \&
  {En{\ss}lin}}]{2004PhRvD..70d3007S}
{Sigl}, G., {Miniati}, F., \& {En{\ss}lin}, T.~A. 2004, Phys. Rev. D, 70,
  043007

\bibitem[{{Stanev}(1997)}]{1997ApJ...479..290S}
{Stanev}, T. 1997, Apj, 479, 290

\bibitem[{{Takeda} {et~al.}(1998){Takeda}, {Hayashida}, {Honda}, {Inoue},
  {Kadota}, {Kakimoto}, {Kamata}, {Kawaguchi}, {Kawasaki}, {Kawasumi},
  {Kitamura}, {Kusano}, {Matsubara}, {Murakami}, {Nagano}, {Nishikawa},
  {Ohoka}, {Sakaki}, {Sasaki}, {Shinozaki}, {Souma}, {Teshima}, {Torii},
  {Tsushima}, {Uchihori}, {Yamamoto}, {Yoshida}, \&
  {Yoshii}}]{1998PhRvL..81.1163T}
{Takeda}, M., {Hayashida}, N., {Honda}, K., {et~al.} 1998, Physical Review
  Letters, 81, 1163

\bibitem[{{The Pierre Auger Collaboration}(2008)}]{2008APh....29..188P}
{The Pierre Auger Collaboration}. 2008, Astroparticle Physics, 29, 188

\bibitem[{{The Pierre Auger collaboration}(2008)}]{2008PhRvL.101f1101A}
{The Pierre Auger collaboration}. 2008, Physical Review Letters, 101, 061101

\bibitem[{{The Pierre Auger Collaboration}(2009)}]{2009arXiv0906.2347T}
{The Pierre Auger Collaboration}. 2009, arxiv:astro-ph/09062347

\bibitem[{{Tinyakov} \& {Tkachev}(2005)}]{2005APh....24...32T}
{Tinyakov}, P.~G. \& {Tkachev}, I.~I. 2005, Astroparticle Physics, 24, 32

\bibitem[{{Toscano} \& {the ANTARES collaboration}(2009)}]{toscano}
{Toscano}, S. \& {the ANTARES collaboration}. 2009, in ICRC 2009 Proceedings

\bibitem[{{V{\'e}ron-Cetty} \& {V{\'e}ron}(2006)}]{2006A&A...455..773V}
{V{\'e}ron-Cetty}, M.-P. \& {V{\'e}ron}, P. 2006, A\&A, 455, 773

\bibitem[{{Vietri}(1995)}]{1995ApJ...453..883V}
{Vietri}, M. 1995, Apj, 453, 883

\bibitem[{{V{\"o}lk} \& {Zirakashvili}(2004)}]{2004A&A...417..807V}
{V{\"o}lk}, H.~J. \& {Zirakashvili}, V.~N. 2004, A\&A, 417, 807

\bibitem[{{Waxman}(1995)}]{1995PhRvL..75..386W}
{Waxman}, E. 1995, Physical Review Letters, 75, 386

\bibitem[{{Zas}(2005)}]{2005NJPh....7..130Z}
{Zas}, E. 2005, New Journal of Physics, 7, 130

\bibitem[{{Zatsepin} \& {Kuzmin}(1966)}]{zatsepin}
{Zatsepin}, Z. \& {Kuzmin}, V. 1966, Zh. Eksp. Teor. Fiz. Pisma Red., 4, 144

\end{thebibliography}


\end{document}